\documentclass[12pt]{article}
\usepackage{natbib}
\usepackage{amsfonts,amsmath,amssymb}
\usepackage{graphicx}
\usepackage{setspace}
\usepackage{fullpage}
\usepackage{authblk}
\usepackage{accents}
\usepackage{xcolor}
\usepackage{multirow}

\newcommand{\iid}{\stackrel{\mathrm{iid}}{\sim}}

\def\citeapos#1{\citeauthor{#1}'s (\citeyear{#1})}
\def\citeApos#1{\citeauthor*{#1}'s (\citeyear{#1})}
\newtheorem{theorem}{Theorem}[section]

\title{Smoothing spline ANOVA for super-large samples: Scalable computation via rounding parameters\footnote{A version of this paper will appear in the upcoming special issue of \emph{Statistics and Its Interface} on Statistical and Computational Theory and Methodology for Big Data.}}
\author[1,2]{Nathaniel E.~Helwig\thanks{helwig@umn.edu}}
\author[3]{Ping Ma\thanks{pingma@uga.edu}}
\affil[1]{\small Department of Psychology, University of Minnesota}
\affil[2]{\small School of Statistics, University of Minnesota}
\affil[3]{\small Department of Statistics, University of Georgia}
\date{July 15, 2015}

\begin{document}

\maketitle

\abstract{
In the current era of big data,  researchers routinely  collect and analyze data of super-large sample sizes. Data-oriented statistical methods have been developed to extract information from super-large data. Smoothing spline ANOVA (SSANOVA) is a promising approach for extracting information from noisy data; however, the heavy computational cost of SSANOVA hinders its wide application. In this paper, we propose a new algorithm for fitting  SSANOVA models to super-large sample data. In this algorithm,  we introduce rounding parameters to make the computation scalable. To demonstrate the benefits of the rounding parameters, we present a simulation study and a real data example using electroencephalography data. Our results reveal that (using the rounding parameters) a researcher can fit nonparametric regression models to very large samples within a few seconds using a standard laptop or tablet computer.
}

\medskip
\emph{Keywords}: Smoothing spline ANOVA, Rounding parameter, Scalable algorithm

\section{Introduction} 
In the current era of big data, it is common for researchers to collect super-large sample data ranging from hundreds of thousands to hundreds of millions of observations. The ambitious BRAIN Initiative of NIH is expected to bring a torrent of data, e.g, 100 terabytes of data per day from a single brain lab.   These super-large datasets provide a wealth of information.  To effectively extract the information, numerous data-oriented statistical learning methods have been developed.  Among these methods,  data-driven nonparametric regression models \citep[see][]{Ruppert+EtAl:2003,Silverman:1985} have achieved remarkable success in identifying subtle patterns and discovering functional relationships in large noisy data; such models require few assumptions about the observed data, but produce a powerful prediction. 

For example, smoothing splines \citep[see][]{Silverman:1985,Wahba:1990} offer a powerful and flexible framework for nonparametric modeling. Smoothing spline analysis of variance (SSANOVA) models \citep{Gu:2013} further expand the research horizon of the smoothing spline; SSANOVAs can model multivariate data and provide nice interpretability of the modeling and prediction outcome. Furthermore, assuming that the smoothing parameters are selected via cross-validation, SSANOVA models have been shown to have desirable asymptotic properties \citep[see][]{Gu:2013,Li:1987,Wahba:1990}. The main drawback of the SSANOVA approach is its computational expense: the computational complexity of SSANOVA is on the order of $O(n^3)$, where $n$ is sample size. 

Over the years, many efforts have been made to design scalable algorithms for SSANOVA.  Generalized additive models \citep[GAMs;][]{Hastie+Tibshirani:1990, Wood:2006} provide scalable computation at the price of eliminating or reparameterizing all interaction terms of an SSANOVA model.  By collapsing similar subspaces, \citet*{Helwig+Ma:2015} provide an algorithm for modeling all interactions with affordable computational complexity. However, even using the most efficient SSANOVA approximation \citep{Kim+Gu:2004,Ma+EtAl:2015} and algorithm \citep{Helwig+Ma:2015}, the computational burden grows linearly with the sample size, which makes the approach impractical for analyzing super-large datasets. 

One possibility is to fit the model to a subset of the observed data. For example, when analyzing ultra large datasets, \citet{Ma+EtAl:2014} suggest fitting regression models to a randomly selected influential sample of the full dataset. This sort of smart-sampling approach works well, as long as a representative sample of observations is selected for analysis; however, the fitted model varies from time to time as the subsample is randomly taken. Furthermore, determining the appropriate size of the subsample could be difficult in some situations.

In this paper, we propose a new approach for fitting SSANOVA models to super-large samples. Specifically, we introduce user-tunable rounding parameters in the SSANOVA model, which makes it possible to control the precision of each predictor. As we demonstrate, fitting a nonparametric regression model to the rounded data can result in substantial computational savings without introducing much bias to the resulting estimate. In the following sections, we provide a brief introduction to SSANOVA (Section~\ref{splines}), develop the concept of rounding parameters for nonparametric regression (Section~\ref{rparms}), present finite-sample and asymptotic results concerning the quality of the rounded SSANOVA estimator (Section~\ref{rqual}), demonstrate the benefits of the rounding parameters with a simulation study (Section~\ref{sim}), and provide an example with real data to reveal the practical potential of the rounding parameters (Section~\ref{ex}).

\section{Smoothing Splines} 
\label{splines}

\subsection{Overview} 
A typical (Gaussian) nonparametric regression model has the form
\begin{equation}
\label{ssa}
y_{i}=\eta(\mathbf{x}_{i})+e_{i}
\end{equation}
where $y_{i}\in\mathbb{R}$ is the response variable, $\mathbf{x}_{i}\equiv(x_{i1},\ldots,x_{ip})$ is the predictor vector, $\eta$ is the unknown smooth function relating the response and predictors, and $e_{i}\iid\mathrm{N}(0,\sigma^{2})$ is unknown, normally-distributed measurement error \citep[see][]{Gu:2013,Ruppert+EtAl:2003,Wahba:1990}. Typically, $\eta$ is estimated by minimizing the penalized least-squares functional
\begin{equation}
\label{penfun}
(1/n)\sum_{i=1}^{n}(y_{i}-\eta(\mathbf{x}_{i}))^{2} + \lambda J(\eta)
\end{equation}
where the nonnegative penalty functional $J$ quantifies the roughness of $\eta$, and the smoothing parameter $\lambda\in(0,\infty)$ balances the trade-off between fitting the data and smoothing $\eta$.

Given fixed smoothing parameters and a set of selected knots $\{\breve{\mathbf{x}}_{h}\}_{h=1}^{q}$, the $\eta_{\lambda}$ minimizing Equation~(\ref{penfun}) can be approximated using
\begin{equation}
\label{ssarep}
\eta_{\lambda}(\mathbf{x}) = \sum_{v=1}^{m}d_{v}\phi_{v}(\mathbf{x}) + \sum_{h=1}^{q}c_{h}\rho_{\mathrm{c}}(\mathbf{x},\breve{\mathbf{x}}_{h})
\end{equation}
where $\{\phi_{v}\}_{v=1}^{m}$ are functions spanning the null space (i.e., $J(\phi_{v})=0$), $\rho_{\mathrm{c}}$ is the reproducing kernel (RK) of the contrast space (i.e., $J(\rho_{\mathrm{c}})>0$), and $\mathbf{d}=\{d_{v}\}_{m\times1}$ and $\mathbf{c}=\{c_{h}\}_{q\times1}$ are the unknown function coefficients \citep[see][]{Helwig+Ma:2015,Kim+Gu:2004,Gu+Wahba:1991}. Note that $\rho_{\mathrm{c}}=\sum_{k=1}^{s}\theta_{k}\rho_{k}^{*}$, where $\rho_{k}^{*}$ denotes the RK of the $k$-th orthogonal contrast space, and $\boldsymbol\theta=(\theta_{1},\ldots,\theta_{s})'$ are additional smoothing parameters with $\theta_{k}\in(0,\infty)$. 

\subsection{Estimation} 

Inserting the optimal representation in Equation~(\ref{ssarep}) into the penalized least-squared functional in Equation~(\ref{penfun}) produces 
\begin{equation}
\label{penloss2}
(1/n)\|\mathbf{y}-\mathbf{K}\mathbf{d}-\mathbf{J}_{\boldsymbol\theta}\mathbf{c}\|^{2} + \lambda\mathbf{c}'\mathbf{Q}_{\boldsymbol\theta}\mathbf{c}
\end{equation}
where $\|\cdot\|^{2}$ denotes the squared Frobenius norm, $\mathbf{y}\equiv\{y_{i}\}_{n\times 1}$, $\mathbf{K}\equiv\{\phi_{v}(\mathbf{x}_{i})\}_{n\times m}$ for $i\in\{1,\ldots,n\}$ and $v\in\{1,\ldots,m\}$, $\mathbf{J}_{\boldsymbol\theta}=\sum_{k=1}^{s}\theta_{k}\mathbf{J}_{k}$ with $\mathbf{J}_{k}\equiv\{\rho_{k}^{*}(\mathbf{x}_{i},\breve{\mathbf{x}}_{h})\}_{n \times q}$ for $i\in\{1,\ldots,n\}$ and $h\in\{1,\ldots,q\}$, and $\mathbf{Q}_{\boldsymbol\theta}=\sum_{k=1}^{s}\theta_{k}\mathbf{Q}_{k}$ where $\mathbf{Q}_{k}\equiv\{\rho_{k}^{*}(\breve{\mathbf{x}}_{g},\breve{\mathbf{x}}_{h})\}_{q \times q}$ for $g,h\in\{1,\ldots,q\}$. Given a choice of $\boldsymbol\lambda\equiv(\lambda/\theta_{1},\ldots,\lambda/\theta_{s})$, the optimal function coefficients are given by
\begin{equation}
\label{coefs}
\begin{split}
\left(\begin{matrix} \hat{\mathbf{d}} \\ \hat{\mathbf{c}} \end{matrix}\right) &=
\left(\begin{matrix} \mathbf{K'K} & \mathbf{K}'\mathbf{J}_{\boldsymbol\theta} \\
 \mathbf{J}_{\boldsymbol\theta}'\mathbf{K} & \mathbf{J}_{\boldsymbol\theta}'\mathbf{J}_{\boldsymbol\theta} + \lambda n\mathbf{Q}_{\boldsymbol\theta} \end{matrix}\right)^{\dagger} \left(\begin{matrix} \mathbf{K}' \\ \mathbf{J}_{\boldsymbol\theta}' \end{matrix}\right)\mathbf{y}
 \end{split}
 \end{equation}
where $(\cdot)^{\dagger}$ denotes the Moore-Penrose pseudoinverse. 

The fitted values are given by $\hat{\mathbf{y}} = \mathbf{K}\hat{\mathbf{d}}+\mathbf{J}_{\boldsymbol\theta}\hat{\mathbf{c}} = \mathbf{S}_{\boldsymbol\lambda}\mathbf{y}$, where 
\begin{equation}
\label{smooth}
\begin{split}
\mathbf{S}_{\boldsymbol\lambda} &= \left(\begin{matrix} \mathbf{K} & \mathbf{J}_{\boldsymbol\theta} \end{matrix}\right)
\left(\begin{matrix} \mathbf{K'K} & \mathbf{K}'\mathbf{J}_{\boldsymbol\theta} \\
 \mathbf{J}_{\boldsymbol\theta}'\mathbf{K} & \mathbf{J}_{\boldsymbol\theta}'\mathbf{J}_{\boldsymbol\theta} + \lambda n\mathbf{Q}_{\boldsymbol\theta} \end{matrix}\right)^{\dagger} \left(\begin{matrix} \mathbf{K}' \\ \mathbf{J}_{\boldsymbol\theta}' \end{matrix}\right)
 \end{split}
 \end{equation}
is the smoothing matrix, which depends on $\boldsymbol\lambda$. The smoothing parameters are typically selected by minimizing \citeApos{Craven+Wahba:1979} generalized cross-validation (GCV) score:
\begin{equation}
\label{gcv}
\mathrm{GCV}(\boldsymbol\lambda) = \{n\|(\mathbf{I}_{n}-\mathbf{S}_{\boldsymbol\lambda})\mathbf{y}\|^{2}\}/\{[n-\mathrm{tr}(\mathbf{S}_{\boldsymbol\lambda})]^{2}\}. 
\end{equation}
The estimates $\hat{\lambda}$ and $\hat{\boldsymbol\theta}$ that minimize the GCV score have desirable properties \citep[see][]{Craven+Wahba:1979,Gu:2013,Gu+Wahba:1991,Li:1987}.

\section{Rounding Parameters}
\label{rparms}

\subsection{Overview}
When fitting a nonparametric regression model to ultra large samples, we propose including user-tunable rounding parameters in the model \citep[see][for preliminary work]{Helwig:Phd}. Assuming that all (continuous) predictors have been transformed to the interval [0,1], the rounding parameters $r_{j}\in(0,1]$ are used to create locally-smoothed versions of the (continuous) predictor variables, such as 
\begin{equation}
\label{round}
z_{ij}=\mathrm{rd}(x_{ij}/r_{j})r_{j}
\end{equation}
for $i\in\{1,\ldots,n\}$ and $j\in\{1,\ldots,p\}$, where the rounding function $\mathrm{rd}(\cdot)$ rounds the input value to the nearest integer. Note that the $z_{ij}$ scores are formed simply by rounding the original $x_{ij}$ scores to the precision defined by the rounding parameter for the $j$-th predictor variable, e.g., if $r_{j}=.02$, then each $x_{ij}$ value is rounded to the nearest .02 to form $z_{ij}$.

Let $\mathbf{z}_{i}\equiv(z_{i1},\ldots,z_{ip})'$ with $z_{ij}$ defined according to Equation~(\ref{round}), and let $\{\breve{\mathbf{z}}_{h}\}_{h=1}^{q}$ denote the rounded knots; then, the penalized least-squares function in Equation~(\ref{penloss2}) can be approximated as $(1/n)\|\mathbf{y}-\mathbf{K}_{\star}\mathbf{d}_{\star}-\mathbf{J}_{\boldsymbol\theta}^{\star}\mathbf{c}_{\star}\|^{2} + \lambda\mathbf{c}_{\star}'\mathbf{Q}^{\star}_{\boldsymbol\theta}\mathbf{c}_{\star}$, where $\mathbf{K}_{\star}$, $\mathbf{J}_{\boldsymbol\theta}^{\star}$, and $\mathbf{Q}^{\star}_{\boldsymbol\theta}$ are defined according to Equation~(\ref{penloss2}) with $\mathbf{z}_{i}$ replacing $\mathbf{x}_{i}$. Similarly, the optimal basis function coefficients corresponding to the rounded data (i.e., $\hat{\mathbf{d}}_{\star}$ and $\hat{\mathbf{c}}_{\star}$) can be defined according to Equation~(\ref{coefs}) with with $\mathbf{z}_{i}$ replacing $\mathbf{x}_{i}$. Finally, smoothing matrix corresponding to these coefficients (denoted by $\mathbf{S}_{\boldsymbol\lambda,r}$) can be defined according to Equation~(\ref{smooth}) with with $\mathbf{z}_{i}$ replacing $\mathbf{x}_{i}$. 

One could calculate the fitted values using $\mathbf{S}_{\boldsymbol\lambda,r}\mathbf{y}$ (and this is what we recommend for the smoothing parameter estimation), however this could introduce a small bias to each predicted score. So, when interpreting specific $\hat{y}_{i}$ scores, we recommend using the $\hat{\mathbf{d}}_{\star}$ and $\hat{\mathbf{c}}_{\star}$ coefficients and basis function matrices with unrounded predictor variable scores
\begin{equation}
\label{fit2}
\begin{split}
\hat{\mathbf{y}}_{\star} &= \mathbf{K}\hat{\mathbf{d}}_{\star} + \mathbf{J}_{\boldsymbol\theta}\hat{\mathbf{c}}_{\star}
 \end{split}
\end{equation}
where $\mathbf{K}$ and $\mathbf{J}_{\boldsymbol\theta}$ are defined according to Equation~(\ref{penloss2}).

\subsection{Computational Benefits}

Let $\{\tilde{\mathbf{z}}_{t}\}_{t=1}^{u}$ denote the set of unique observed $\mathbf{z}_{i}$ vectors with $u \geq q$, and note that $u$ has an upper-bound that is determined by the rounding parameters and the predictor variables. For example, suppose that $\tilde{\mathbf{z}}_{t}\equiv(\tilde{z}_{t1},\tilde{z}_{t2})$ with $\tilde{z}_{t1}\in[0,1]$ and $\tilde{z}_{t2}\in\{1,\ldots,f\}$; then, defining $r_{1}=.01$, it is evident that $u \leq 101f$, given that $z_{ij}$ can have a maximum of 101 unique values for the first predictor, and maximum of $f$ unique values for the second predictor. As a second example, suppose that $\tilde{\mathbf{z}}_{t}\equiv(\tilde{z}_{t1},\tilde{z}_{t2})$ with $\tilde{z}_{t1},\tilde{z}_{t2}\in[0,1]$; then, defining $r_{1}=r_{2}=.01$, it is evident that $u \leq 101^{2}$, given that $z_{ij}$ can have a maximum of 101 unique values for each predictor. Similar reasoning can be used to place an upper bound on $u$ for different combinations of rounding parameters and predictor variable types. 

Note that the inner-portion of $\mathbf{S}_{\boldsymbol\lambda,r}$ can be written as
\begin{equation}
\begin{split}
\label{usmooth}
&\left(\begin{matrix} \mathbf{K}_{\star}'\mathbf{K}_{\star} & \mathbf{K}_{\star}'\mathbf{J}_{\boldsymbol\theta}^{\star} \\
 (\mathbf{J}_{\boldsymbol\theta}^{\star})'\mathbf{K}_{\star} & (\mathbf{J}_{\boldsymbol\theta}^{\star})'\mathbf{J}_{\boldsymbol\theta}^{\star} + \lambda n \mathbf{Q}^{\star}_{\boldsymbol\theta} \end{matrix}\right)^{\dagger}  = \\ 
 & \qquad \left(\begin{matrix} \tilde{\mathbf{K}}_{\star}'\mathbf{W}\tilde{\mathbf{K}}_{\star} & \tilde{\mathbf{K}}_{\star}'\mathbf{W}\tilde{\mathbf{J}}^{\star}_{\boldsymbol\theta} \\
 (\tilde{\mathbf{J}}_{\boldsymbol\theta}^{\star})'\mathbf{W}\tilde{\mathbf{K}}_{\star} & (\tilde{\mathbf{J}}^{\star}_{\boldsymbol\theta})'\mathbf{W}\tilde{\mathbf{J}}^{\star}_{\boldsymbol\theta} + \lambda n\mathbf{Q}^{\star}_{\boldsymbol\theta} \end{matrix}\right)^{\dagger}
 \end{split}
\end{equation}
where $\tilde{\mathbf{K}}_{\star}\equiv\{\phi_{v}(\tilde{\mathbf{z}}_{t})\}_{u\times m}$ for $t\in\{1,\ldots,u\}$ and $v\in\{1,\ldots,m\}$, $\tilde{\mathbf{J}}^{\star}_{\boldsymbol\theta}=\sum_{k=1}^{s}\theta_{k}\tilde{\mathbf{J}}^{\star}_{k}$ where $\tilde{\mathbf{J}}^{\star}_{k}\equiv\{\rho_{k}^{*}(\tilde{\mathbf{z}}_{t},\breve{\mathbf{z}}_{h})\}_{u \times q}$ for $t\in\{1,\ldots,u\}$ and $h\in\{1,\ldots,q\}$, and $\mathbf{W}\equiv\mathrm{diag}(w_{1},\ldots,w_{u})$ with $w_{t}$ denoting the number of $\mathbf{z}_{i}$ that are equal to $\tilde{\mathbf{z}}_{t}$ (for $t\in\{1,\ldots,u\}$). Next, define $\tilde{\mathbf{X}}=(\tilde{\mathbf{K}}_{\star}, \tilde{\mathbf{J}}^{\star}_{\boldsymbol\theta})$ and define the reduced smoothing matrix $\tilde{\mathbf{S}}_{\boldsymbol\lambda}^{\star}$, such as
 \begin{equation}
\label{rsmooth}
\tilde{\mathbf{S}}_{\boldsymbol\lambda}^{\star} = \tilde{\mathbf{X}}
\left(\begin{matrix} \tilde{\mathbf{K}}_{\star}'\mathbf{W}\tilde{\mathbf{K}}_{\star} & \tilde{\mathbf{K}}_{\star}'\mathbf{W}\tilde{\mathbf{J}}^{\star}_{\boldsymbol\theta} \\
 (\tilde{\mathbf{J}}_{\boldsymbol\theta}^{\star})'\mathbf{W}\tilde{\mathbf{K}}_{\star} & (\tilde{\mathbf{J}}^{\star}_{\boldsymbol\theta})'\mathbf{W}\tilde{\mathbf{J}}^{\star}_{\boldsymbol\theta} + \lambda n\mathbf{Q}^{\star}_{\boldsymbol\theta} \end{matrix}\right)^{\dagger} \tilde{\mathbf{X}}'.
\end{equation}
Note that  $\tilde{\mathbf{S}}_{\boldsymbol\lambda}^{\star}$ is a  $u \times u$ matrix, and note that $u<n$ if there are replicate predictor vectors after the rounding (which is guaranteed if $n$ is larger than $u$'s upper bound). 

Next, suppose that the $(y_{i},\mathbf{z}_{i})$ scores are ordered such that observations $1,\ldots,w_{1}$ have predictor scores $\tilde{\mathbf{z}}_{1}$, observations $w_{1}+1,\ldots,w_{1}+w_{2}$ have predictor scores $\tilde{\mathbf{z}}_{2}$, and so on. Then $\mathbf{S}_{\boldsymbol\lambda,r}$ can be written in terms of $\tilde{\mathbf{S}}_{\boldsymbol\lambda}^{\star}$, such as
\begin{equation}
\label{r2fsmooth}
\mathbf{S}_{\boldsymbol\lambda,r} = \left(\begin{matrix}
(\mathbf{e}_{1}'\tilde{\mathbf{S}}_{\boldsymbol\lambda}^{\star}\mathbf{e}_{1})\mathbf{1}_{w_{1}}\mathbf{1}_{w_{1}}' & \cdots & (\mathbf{e}_{1}'\tilde{\mathbf{S}}_{\boldsymbol\lambda}^{\star}\mathbf{e}_{u})\mathbf{1}_{w_{1}}\mathbf{1}_{w_{u}}'  \\
(\mathbf{e}_{2}'\tilde{\mathbf{S}}_{\boldsymbol\lambda}^{\star}\mathbf{e}_{1})\mathbf{1}_{w_{2}}\mathbf{1}_{w_{1}}' &  \cdots & (\mathbf{e}_{2}'\tilde{\mathbf{S}}_{\boldsymbol\lambda}^{\star}\mathbf{e}_{u})\mathbf{1}_{w_{2}}\mathbf{1}_{w_{u}}'  \\
\vdots & \ddots & \vdots \\
(\mathbf{e}_{u}'\tilde{\mathbf{S}}_{\boldsymbol\lambda}^{\star}\mathbf{e}_{1})\mathbf{1}_{w_{u}}\mathbf{1}_{w_{1}}' & \cdots & (\mathbf{e}_{u}'\tilde{\mathbf{S}}_{\boldsymbol\lambda}^{\star}\mathbf{e}_{u})\mathbf{1}_{w_{u}}\mathbf{1}_{w_{u}}'  \\
\end{matrix}\right)
\end{equation}
where $\mathbf{e}_{t}$ denotes a $u \times 1$ vector with a one in the $t$-th position and zeros elsewhere, and $\mathbf{1}_{w_{t}}$ denotes a $w_{t} \times 1$ vector of ones (for $t\in\{1,\ldots,u\}$). Furthermore, note that the fitted values corresponding to $\mathbf{S}_{\boldsymbol\lambda,r}$ can be written as 
\begin{equation}
\begin{split}
\label{fit4}
\mathbf{S}_{\boldsymbol\lambda,r}\mathbf{y} = \left(\begin{matrix}(\mathbf{e}_{1}'\tilde{\mathbf{S}}_{\boldsymbol\lambda}^{\star}\mathbf{e}_{1})\mathbf{1}_{w_{1}} & \cdots & (\mathbf{e}_{1}'\tilde{\mathbf{S}}_{\boldsymbol\lambda}^{\star}\mathbf{e}_{u})\mathbf{1}_{w_{1}}  \\
(\mathbf{e}_{2}'\tilde{\mathbf{S}}_{\boldsymbol\lambda}^{\star}\mathbf{e}_{1})\mathbf{1}_{w_{2}} & \cdots & (\mathbf{e}_{2}'\tilde{\mathbf{S}}_{\boldsymbol\lambda}^{\star}\mathbf{e}_{u})\mathbf{1}_{w_{2}}  \\
\vdots & \ddots & \vdots \\
(\mathbf{e}_{u}'\tilde{\mathbf{S}}_{\boldsymbol\lambda}^{\star}\mathbf{e}_{1})\mathbf{1}_{w_{u}} & \cdots & (\mathbf{e}_{u}'\tilde{\mathbf{S}}_{\boldsymbol\lambda}^{\star}\mathbf{e}_{u})\mathbf{1}_{w_{u}}  \\
\end{matrix}\right)\tilde{\mathbf{y}}
\end{split}
\end{equation}
where $\tilde{\mathbf{y}}\equiv\{\tilde{y}_{t}\}_{u \times 1}$ with $\tilde{y}_{t}=\sum_{\mathcal{I}_{t}}y_{i}$ and $\mathcal{I}_{t}\subset\{1,\ldots,n\}$ denoting the set of indices such that $\mathbf{z}_{i}$ is equal to $\tilde{\mathbf{z}}_{t}$.

Now, let $\hat{y}_{t}^{\star} = \mathbf{e}_{t}'\tilde{\mathbf{S}}_{\boldsymbol\lambda}^{\star}\tilde{\mathbf{y}}$ denote the fitted value corresponding to $\tilde{\mathbf{z}}_{t}$ (for $t\in\{1,\ldots,u\}$), and note that the numerator of the GCV score in Equation~(\ref{gcv}) can be written as
\begin{equation}
\label{GCV1}
\begin{split}
n\sum_{t=1}^{u}\sum_{\mathcal{I}_{t}}(y_{i}-\hat{y}_{t}^{\star})^{2} &= n\sum_{i=1}^{n}y_{i}^{2} - 2n\sum_{t=1}^{u}\tilde{y}_{t}\hat{y}_{t}^{\star} + n\sum_{t=1}^{u}w_{t}(\hat{y}_{t}^{\star})^{2}\\
&= n\left[\|\mathbf{y}\|^{2} - 2\tilde{\mathbf{y}}'\tilde{\mathbf{S}}_{\boldsymbol\lambda}^{\star}\tilde{\mathbf{y}} + \tilde{\mathbf{y}}'\tilde{\mathbf{S}}_{\boldsymbol\lambda}^{\star}\mathbf{W}\tilde{\mathbf{S}}_{\boldsymbol\lambda}^{\star}\tilde{\mathbf{y}}\right]
\end{split}
\end{equation}
In addition, note that the denominator of the GCV score can be written as $[n-\mathrm{tr}(\mathbf{S}_{\boldsymbol\lambda,r})]^{2} = [n - \mathrm{tr}(\mathbf{W}\tilde{\mathbf{S}}_{\boldsymbol\lambda}^{\star})]^{2}$ using the relation in Equation~(\ref{r2fsmooth}). 

The above formulas imply that, after initializing $\tilde{\mathbf{y}}$, $\|\mathbf{y}\|^{2}$, and $\mathbf{W}$, it is only necessary to calculate the reduced smoothing matrix $\tilde{\mathbf{S}}_{\boldsymbol\lambda}^{\star}$ to evaluate the GCV score. Furthermore, note that the optimal function coefficients can be estimated from the reduced smoothing matrix using
\begin{equation}
\label{cdhatROUND2}
\left(\begin{matrix} \hat{\mathbf{d}}_{\star} \\ \hat{\mathbf{c}}_{\star} \end{matrix}\right) = \left(\begin{matrix} \tilde{\mathbf{K}}_{\star}'\mathbf{W}\tilde{\mathbf{K}}_{\star} & \tilde{\mathbf{K}}_{\star}'\mathbf{W}\tilde{\mathbf{J}}^{\star}_{\boldsymbol\theta} \\
 (\tilde{\mathbf{J}}_{\boldsymbol\theta}^{\star})'\mathbf{W}\tilde{\mathbf{K}}_{\star} & (\tilde{\mathbf{J}}^{\star}_{\boldsymbol\theta})'\mathbf{W}\tilde{\mathbf{J}}^{\star}_{\boldsymbol\theta} + \lambda n\mathbf{Q}^{\star}_{\boldsymbol\theta} \end{matrix}\right)^{\dagger} \left(\begin{matrix}
\tilde{\mathbf{K}}_{\star}'\\ 
(\tilde{\mathbf{J}}^{\star}_{\boldsymbol\theta})' \end{matrix}\right) \tilde{\mathbf{y}}
\end{equation}
which implies that it is never necessary to construct the full $n \times n$ smoothing matrix to estimate $\eta$ when using the rounding parameters.

\subsection{Choosing Rounding Parameters}
In many situations, a rounding parameter can be determined by the measurement precision of the predictor variable. For example, suppose we have one predictor $x_{i}$ recorded with the precision of two decimals on the interval [0,1], i.e., $x_{i}\in\{0,0.01,0.02,\ldots,0.99,1\}$ for $i\in\{1,\ldots,n\}$. In this case, setting $r=0.01$ will produce the exact same solution as using the unrounded predictors (i.e., $z_{i} = x_{i} \forall i$) and can immensely reduce the computational burden. Note that $u \leq 101$ even if $n$ is very large, and it is only necessary to evaluate the functions $\{\phi_{v}\}_{v=1}^{m}$ and $\rho_{\mathrm{c}}$ for the $u \ll n $ unique predictor scores to estimate $\eta$.

\begin{figure*}
\centering
 \scalebox{.5}{\includegraphics{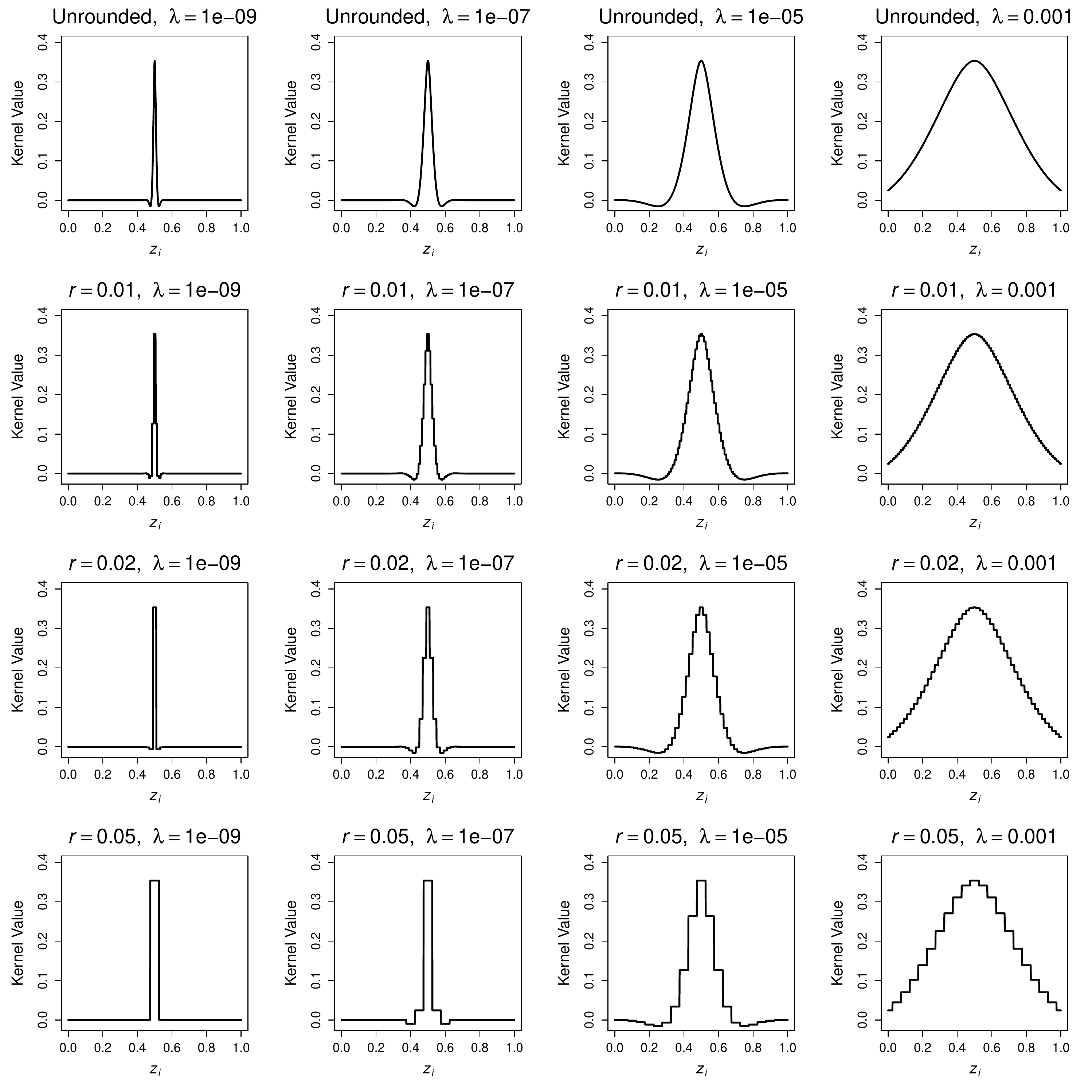}} 
 \caption{\label{fig1} Asymptotic cubic spline kernel function for $z_{i}\in[0,1]$ and $\breve{z}=0.5$.}
\end{figure*}

Now, for large $n$, note that a cubic smoothing spline is approximately a weighted moving average smoother \citep[see][Section~3]{Silverman:1985}. In particular, let $s_{i_{1}i_{2}(\lambda)}$ denote the entry in the $i_{1}$-th row and $i_{2}$-th column of $\mathbf{S}_{\lambda}$, and note that $s_{i_{1}i_{2}(\lambda)}$ asymptotically depends on a kernel function whose influence decreases exponentially as $|x_{i_{1}}-x_{i_{2}}|$ increases \citep[see][equations~3.1--3.4]{Silverman:1985}. Also, note that the rounding parameter proposed in this paper widens the peak of the kernel (see Figure~\ref{fig1}).
For relatively smooth functions (e.g., $\lambda \geq 10^{-3}$), the shape of the asymptotic kernel function is stable for $r\leq0.05$; however, for more jagged functions (e.g., $\lambda \leq10^{-7}$), the rounding parameter will need to be set smaller (e.g., $r=0.01$) for the rounded kernel function to resemble the true asymptotic kernel (see Figure~\ref{fig1}).

\section{Quality of Rounded Solution}
\label{rqual}

\subsection{A Taylor Heuristic}

Note that the rounded predictor $z_{ij}$ can be written as
\begin{equation}
z_{ij} = x_{ij} + r_{j}v_{ij}
\end{equation}
where $v_{ij} = (z_{ij}-x_{ij})/r_{j}$ by definition and $|z_{ij}-x_{ij}| \leq r_{j}/2$ so that $|v_{ij}| \leq 1/2$. This implies $
\mathbf{z}_{i} = \mathbf{x}_{i} + \mathbf{R}\mathbf{v}_{i}$  where  $\mathbf{v}_{i}=(v_{i1},\ldots,v_{ip})'$ and $\mathbf{R} = \mathrm{diag}(r_{1},\ldots,r_{p})$. Consider the linear approximation of $\eta(\mathbf{z}_{i})$ at the point $\mathbf{x}_{i}$
\[
\eta(\mathbf{z}_{i}) = \eta(\mathbf{x}_{i}) + [\nabla \eta(\mathbf{x}_{i})]' \mathbf{R}\mathbf{v}_{i} + o(\|\mathbf{R}\mathbf{v}_{i}\|)
\]
where $\nabla \eta$ denotes the gradient  of $\eta$. If the gradient of $\eta$ were known, we could approximate the rounding error using
\begin{equation}
\begin{split}
n^{-1}\sum_{i=1}^{n}[\eta(\mathbf{x}_{i}) - \eta(\mathbf{z}_{i})]^{2} &\approx n^{-1}\sum_{i=1}^{n} \{ [\nabla \eta(\mathbf{x}_{i})]' \mathbf{R}\mathbf{v}_{i} \}^{2}\\
& \leq  n^{-1}\sum_{i=1}^{n} \| \nabla \eta (\mathbf{x}_{i}) \|^{2} \|\mathbf{R}\mathbf{v}_{i}\|^{2}\\
& \leq  (4n)^{-1}\sum_{i=1}^{n} \| \nabla \eta(\mathbf{x}_{i}) \|^{2} \|\mathbf{r}\|^{2}
\end{split}
\end{equation}
where $\mathbf{r}=(r_{1},\ldots,r_{p})'$; note that the last line is due to the fact that $|v_{ij}| \leq 1/2$.

For example, using an $m$-th order polynomial smoothing spline with $x_{i}\in[0,1]$ \citep[see][]{Craven+Wahba:1979,Gu:2013} we have
\[
\eta_{\lambda}(x) = \sum_{v=0}^{m-1}d_{v}k_{v}(x) + \sum_{h=1}^{q}c_{h} \rho_{\breve{x}_{h}}(x)
\]
where $k_{v}(\cdot)$ are scaled Bernoulli polynomials, $\{\breve{x}_{h}\}_{h=1}^{q} \subset \{x_{i}\}_{i=1}^{n}$ are the selected knots, and
\[
\rho_{\breve{x}_{h}}(x) = k_{m}(x)k_{m}(\breve{x}_{h}) + (-1)^{m-1}k_{2m}(|x-\breve{x}_{h}|)
\]
is the reproducing kernel of the contrast space. Using the properties of Bernoulli polynomials we have
\[
\eta_{\lambda}'(x) = \frac{\partial \eta_{\lambda}(x)}{\partial x} = \sum_{v=1}^{m-1}d_{v}k_{v-1}(x) + \sum_{h=1}^{q}c_{h} \rho_{\breve{x}_{h}}'(x)
\]
where 
\[
\rho_{\breve{x}_{h}}'(x) = k_{m-1}(x)k_{m}(\breve{x}_{h}) + (-1)^{m-1}s_{h}k_{2m-1}(x-\breve{x}_{h})
\]
with $s_{h}=1$ if $x\geq \breve{x}_{h}$ and $s_{h}=-1$ otherwise \citep[see][]{Craven+Wahba:1979,Gu:2013}. 

\begin{figure*} 
\centering
 \scalebox{.7}{\includegraphics{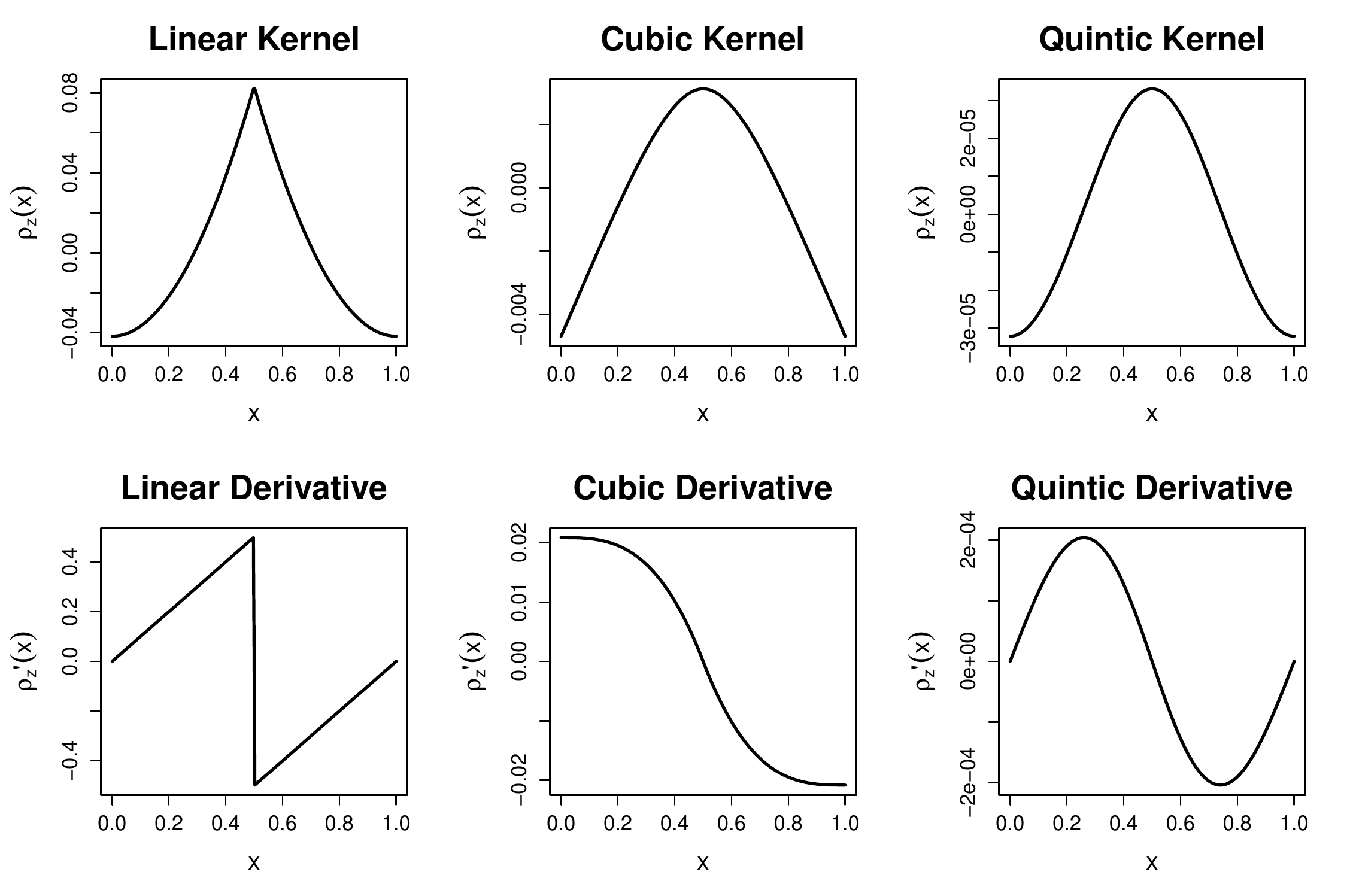}} 
 \caption{\label{fig2} Top: contrast reproducing kernel $\rho_{z}(x)$ for linear spline ($m=1$), cubic spline ($m=2$), and quintic spline ($m=3$) with $z=0.5$ as the knot. Bottom: contrast reproducing kernel derivative $\rho_{z}'(x)$ for $m$-th order polynomial splines.}
\end{figure*}

Consequently, for polynomial splines we can approximate the rounding error using
\[
\begin{split}
n^{-1}\sum_{i=1}^{n}[\eta(x_{i}) - \eta(z_{i})]^{2} &\approx n^{-1}\sum_{i=1}^{n}(rv_{i})^{2} [\eta_{\lambda}'(x_{i})]^{2}\\
& \leq r^{2}(4n)^{-1} \|\mathbf{X}\mathbf{b}\|^{2}\\
\end{split}
\]
where $\mathbf{X} = [\tilde{\mathbf{K}},\tilde{\mathbf{J}}]$ with $\tilde{\mathbf{K}} = \{k_{v}(x_{i})\}_{n \times m-1}$ for $v\in\{0,\ldots,m-2\}$ and $\tilde{\mathbf{J}} = \{\rho_{\breve{x}_{h}}'(x_{i})\}_{n \times q}$ for $h\in\{1,\ldots,q\}$, and $\mathbf{b}=(d_{1},\ldots,d_{m-1}, c_{1},\ldots,c_{q} )'$. Note that the contrast space reproducing kernel $\rho_{\breve{x}_{h}}(x)$ is rather smooth for the classic cubic smoothing spline, and the magnitude of the derivatives are rather small (see Figure~\ref{fig2}). This implies that setting $r\in\{0.01,0.02,0.05\}$ will not introduce much rounding error to the contrast kernel evaluation when using cubic smoothing splines on $x_{i}\in[0,1]$.

The rounding error depends on the norm $\|\mathbf{X}\mathbf{b}\|$, so the relative impact of a particular choice of rounding parameters will depend on the (unknown) function coefficients $\mathbf{b}$. For practical use, we can approximate the rounding error relative to the norm of the coefficients, such as
\[
\begin{split}
\frac{1}{n\|\mathbf{b}\|^{2}}\sum_{i=1}^{n}[\eta(x_{i}) - \eta(z_{i})]^{2} &\approx \frac{1}{n\|\mathbf{b}\|^{2}}\sum_{i=1}^{n}(rv_{i})^{2} [\eta_{\lambda}'(x_{i})]^{2}\\
& \leq r^{2}(4n)^{-1} \lambda_{1}^{*} \\
\end{split}
\]
where $\lambda_{1}^{*}$ is the largest eigenvalue of $\mathbf{X}'\mathbf{X}$; note that we have $\|\mathbf{X}\mathbf{b}\|^{2} \leq \|\mathbf{X}\|^{2} \|\mathbf{b}\|^{2}$ and  $\|\mathbf{X}\|^{2}=\lambda_{1}^{*}$ by definition. For practical computation, it is possible to estimate $\lambda_{1}^{*}/n$ by taking a random sample of $\tilde{n} \ll n$ observations, and then approximate the relative rounding error as $r^{2}(\tilde{n}4)^{-1} \hat{\lambda}_{1}^{*}$. Clearly this sort of approach can be extended to assess the relative rounding error for tensor product smoothing splines, but the gradient formulas become a bit more complicated.

\subsection{Finite Sample Performance}

To quantify the finite-sample error introduced by rounding, define the loss function
\begin{equation}
\begin{split}
L(r) &= \frac{1}{n}\sum_{i=1}^{n} \left( \hat{\eta}_{\lambda}(\mathbf{x}_{i}) - \hat{\eta}_{\lambda,r}(\mathbf{z}_{i}) \right)^2 \\
&= n^{-1} \| (\mathbf{S}_{\boldsymbol\lambda} - \mathbf{S}_{\boldsymbol\lambda,r})\mathbf{y} \|^{2} 
\end{split}
\end{equation}
where $\mathbf{S}_{\boldsymbol\lambda}$ and $\mathbf{S}_{\boldsymbol\lambda,r}$ are the smoothing matrices corresponding to the unrounded and rounded predictors (i.e., $\mathbf{x}_{i}$ and $\mathbf{z}_{i}$, respectively). Denote the risk function as
\begin{equation}
\begin{split}
R(r) &= E[L(r)]\\
&= n^{-1} \| (\mathbf{S}_{\boldsymbol\lambda} - \mathbf{S}_{\boldsymbol\lambda,r})\boldsymbol\eta \|^{2} + n^{-1}\sigma^{2}\mathrm{tr}\{(\mathbf{S}_{\boldsymbol\lambda} - \mathbf{S}_{\boldsymbol\lambda,r})^{2}\}
\end{split}
\end{equation}
where $\boldsymbol\eta=\{\eta(\mathbf{x}_{i})\}_{n\times1}$ contains the realizations of the (unknown) true function $\eta$. Note that the first term of $R(r)$ corresponds to the (squared) bias difference between $\hat{\eta}_{\lambda}$ and $\hat{\eta}_{\lambda,r}$, and the second term is related to (but not equal to) the variance difference. Also note that we can write
\begin{equation}
\begin{split}
R(r) &\leq n^{-1} \|\mathbf{S}_{\boldsymbol\lambda} - \mathbf{S}_{\boldsymbol\lambda,r} \|^{2} \| \boldsymbol\eta \|^{2} + n^{-1}\sigma^{2}\sum_{i=1}^{n}\lambda_{i,r}\\
& \leq \lambda_{1,r} \left( n^{-1} \| \boldsymbol\eta \|^{2} + \sigma^{2} \right)
\end{split}
\end{equation}
where $\lambda_{1,r} \geq \cdots \geq \lambda_{n,r}$ are the eigenvalues of $(\mathbf{S}_{\boldsymbol\lambda} - \mathbf{S}_{\boldsymbol\lambda,r})^{2}$. 

The risk $R(r)$ depends on the squared norm of the unknown function $\eta$, so the practical relevance of a particular value of $R(r)$, e.g., $R(r) = 0.1$, differs depending on the situation, i.e., unknown true function. To overcome this practical issue, we can examine the risk relative to the squared norm of the unknown function, such as
\begin{equation}
\label{relrisk}
\begin{split}
U(r) &= R(r)\| \boldsymbol\eta \|^{-2}\\
&\leq n^{-1} \lambda_{1,r} \left( 1 + n\sigma^{2}\| \boldsymbol\eta \|^{-2} \right)
\end{split}
\end{equation}
where $n\sigma^{2}\| \boldsymbol\eta \|^{-2} = \sigma^{2}/(\|\boldsymbol\eta \|^{2}/n)$ relates to the noise-to-signal ratio, i.e., inverse of signal-to-noise ratio (SNR). Furthermore, for a fixed SNR and a large enough $n$, the second term in the upper-bound of the relative risk is negligible, and we have that  $U(r) \lesssim n^{-1} \lambda_{1,r}$. Consequently, it is only necessary to know the largest eigenvalue of $\mathbf{S}_{\boldsymbol\lambda} - \mathbf{S}_{\boldsymbol\lambda,r}$ to understand the expected performance of a given set of rounding parameters for a large sample size $n$.

In practice, calculating $\mathbf{S}_{\boldsymbol\lambda} - \mathbf{S}_{\boldsymbol\lambda,r}$ and $\lambda_{1,r}$ for various values of $r$ is a computational challenge for large $n$. For practical computation, we recommend examining $R(r)$ and/or $U(r)$ using a random sample of $\tilde{n} \ll n $ observations. Using this approach, the unknown parameters (i.e., $\boldsymbol\eta$ and $\sigma^{2})$ can be estimated using the results of the unrounded solution. For example, the SNR can be estimated as $(\| \hat{\boldsymbol\eta} \|^{2}/\tilde{n})/\hat{\sigma}^{2}$ where $\hat{\boldsymbol\eta}$ and $\hat{\sigma}^{2}$ are the estimated function and error variance using the $\tilde{n}$ observations with unrounded predictors. Or, if the approximate SNR is known, Equation~(\ref{relrisk}) can be used to place an upper-bound on the relative risk $U(r)$.

\begin{figure*} 
\centering
 \scalebox{.65}{\includegraphics{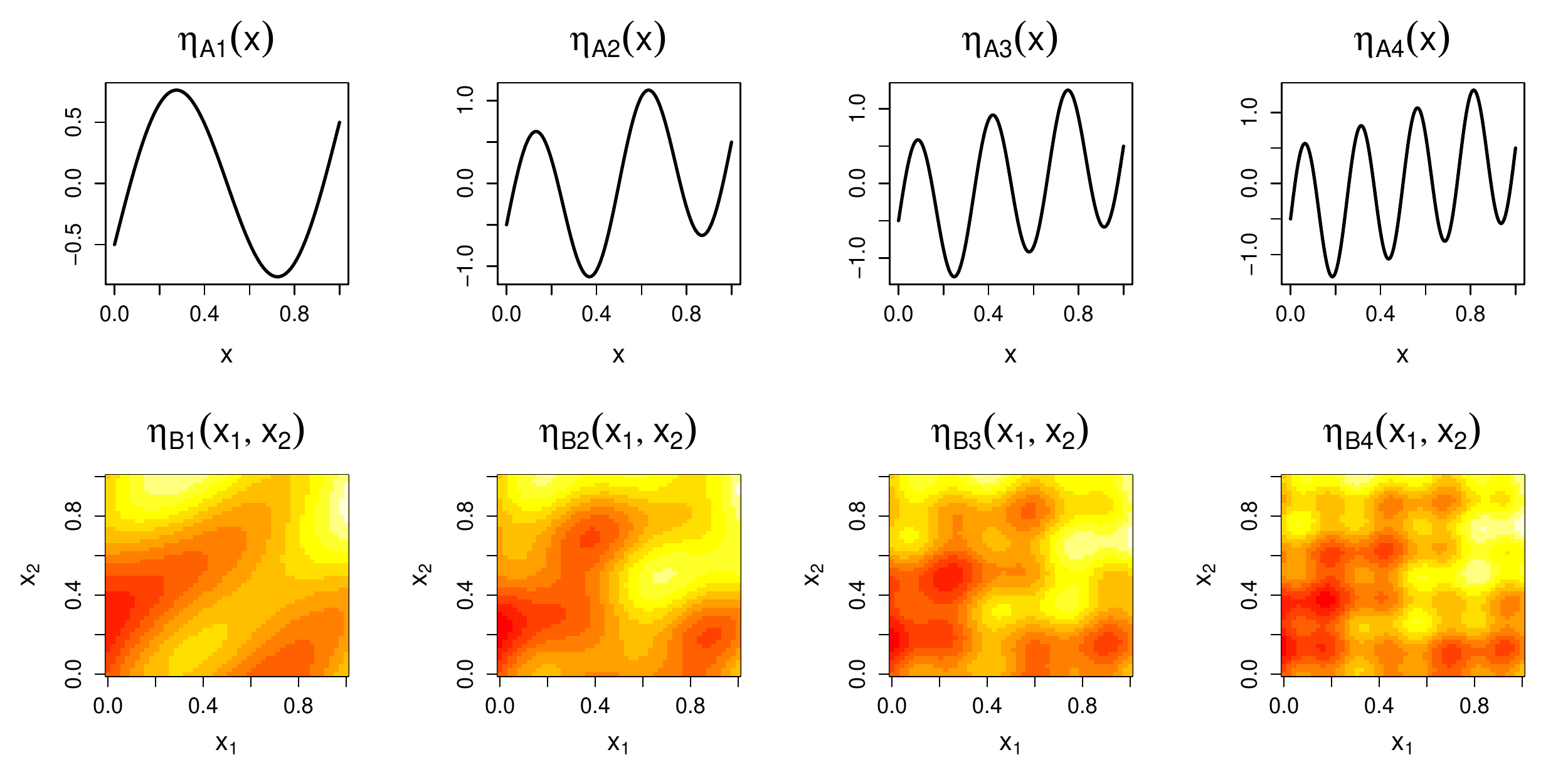}} 
 \caption{\label{fig3} Functions with various degrees of smoothness. $\eta_{Ak}(x) = x - 0.5 + \sin(2 k \pi x)$ for $x\in[0,1]$ and $\eta_{Bk}(x_{1},x_{2}) = x_{1} + x_{2} - 1 + [ \sin(2 k \pi x_{1}) + \cos(2 k \pi x_{2}) + 2\sin(2 \pi (x_{1}-x_{2}) ) ]/4$ for $x_{1},x_{2}\in[0,1]$}
\end{figure*}

\begin{figure*} 
\centering
 \scalebox{.65}{\includegraphics{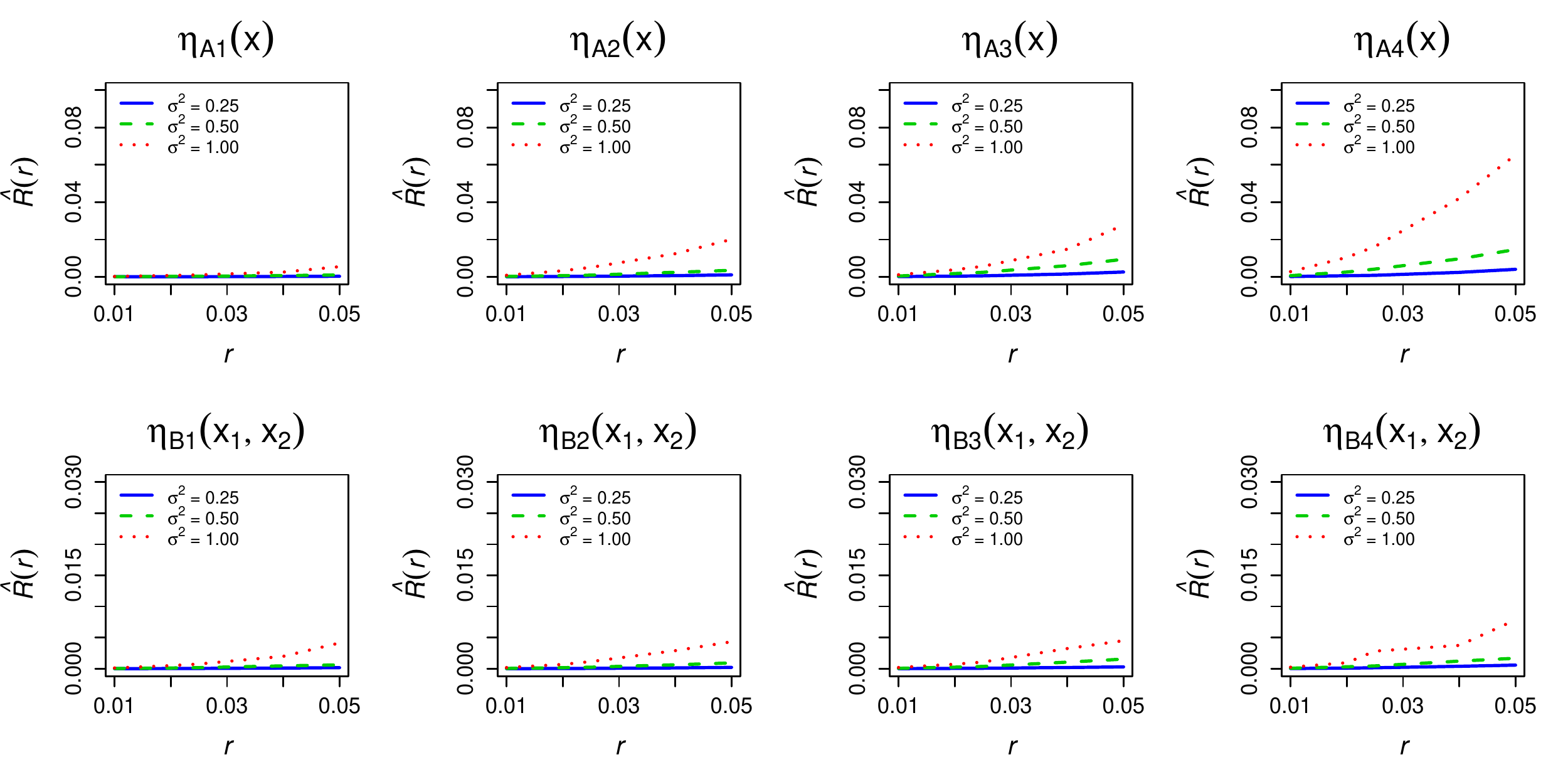}} 
 \caption{\label{fig4} Median estimated risk $\hat{R}(r) = \tilde{n}^{-1} \| (\mathbf{S}_{\boldsymbol\lambda} - \mathbf{S}_{\boldsymbol\lambda,r})\hat{\boldsymbol\eta} \|^{2} + \tilde{n}^{-1}\hat{\sigma}^{2}\mathrm{tr}\{(\mathbf{S}_{\boldsymbol\lambda} - \mathbf{S}_{\boldsymbol\lambda,r})^{2}\}$ for various functions, rounding parameters, and error variances using five random samples of $\tilde{n}=500$ observations.}
\end{figure*}

We demonstrate this approach in Figures~\ref{fig3}--\ref{fig4}, which plot functions with various degrees of smoothness (Figure~\ref{fig3}) and the median estimated rounding risk $\hat{R}(r)$ across five samples of $\tilde{n}=500$ observations (Figure~\ref{fig4}). Note that Figure~\ref{fig4} illustrates that the expected difference between the unrounded and rounded solutions increases as the error variance increases. Furthermore, note that Figure~\ref{fig4} affirms that for $x\in[0,1]$ setting $r=0.01$ can be expected to introduce minimal rounding error for a variety of functions and SNRs. Finally, Figure~\ref{fig4} reveals that setting $r\in\{0.01,0.02,0.05\}$ will not introduce much rounding error whenever the underlying function $\eta$ is relatively smooth. For example, for the functions $\eta_{A1}$ and $\eta_{B1}$, we should expect a negligible difference between the unrounded and rounded solutions using $r=0.05$ for a variety of different SNRs.

\subsection{Asymptotic Bias and Variance}

To establish the asymptotic properties of the proposed estimate, we employ an equivalent kernel approach developed in \cite{Nychka:1995}. The key idea is that a smoothing spline estimate can be written as kernel estimate
\begin{equation}
\hat{\eta}_{\lambda}(x)= \frac{1}{n}\sum_{i=1}^{n}w(x_i, x)y_i
\end{equation}
where the kernel function $w(x_i, x)$ can be well approximated by a Green's function. Then the asymptotic properties of $\hat{\eta}_{\lambda}$ can be established via the analytical properties of the Green's function.

Following \cite{Nychka:1995}, we establish the asymptotic properties of our rounding estimate for the one dimensional case.
In addition, we assume that we use a full basis where all distinct rounded data are used as knots, i.e., $q=u$. Then our estimate $\hat{\eta}_{\lambda,r}$ is the minimizer of 
\begin{equation}
\label{penfunround}
(1/n)\sum_{i=1}^{n}(y_{i}-\eta({z}_{i}))^{2} + \lambda \int_0^1 (\eta^{(m)})^2dx.
\end{equation}

Let $F_{n,r}$ denote the empirical distribution function for the rounded predictor $z_i$, $ i=1, \ldots, n$, let $F$ be the limiting distribution of the original predictor $x$ with a continuous and strictly positive density function $f$ on $[0, 1]$ and let
\[
D_{n,r} = \sup_{x \in [0,1]}| F_{n,r} -F |,
\]
and $\rho=\lambda^{1/2m}$. Then we have the following theorem.
\begin{theorem}
Assume that $\hat{\eta}_{\lambda,r}$ is a smoothing spline estimate of (\ref{penfunround}) with $m=1$ and $z_i$ are not equally spaced. Suppose that $\eta \in C_2[0,1]$ and satisfies the H\"{o}lder condition 
$|\eta^{(2)}(x)-\eta^{(2)}(x^{\prime})| \le M |x-x^{\prime}|^{\beta}$ for some $\beta > 0$ and some $M < \infty$. Assume that $f$ has a uniformly continuous derivative and $D_{n,r}  \rightarrow 0$ as $n \rightarrow \infty$.
Choose $0 < \Delta < 1$ and let  $\lambda_n  \rightarrow 0$ and $\Lambda_n  \rightarrow 0$ as $ n \rightarrow \infty$. Then
\[
\begin{split}
E[\hat{\eta}_{\lambda,r}(x)]- \eta(x) &= -\frac{\lambda}{f(x)}\eta^{(2)}(x) +o(\lambda) + O(\frac{D_{n,r}}{\rho}),\\
\text{Var}[\hat{\eta}_{\lambda,r}(x)] &= \frac{\sigma^2}{8nf(x)} (\frac{ f(x)}{\lambda} )^{1/2} + \sigma^2 O(\frac{D_{n,r}}{\rho}),
\end{split}
\]
uniformly for $\lambda \in [\lambda_n, \Lambda_n]$ and $x \in [\Delta, 1-\Delta]$ as $ n \rightarrow \infty$.
\end{theorem}
\noindent The theorem is a direct result of Theorem 2.2 of  \cite{Nychka:1995}. For $m > 1$, a slightly more complicated version of our theorem can be shown using Theorem 2  of \cite{Wang+EtAl:2013}. 

The theorem states that both the bias and variance of our estimate $\hat{\eta}_{\lambda,r}$ depend on $D_{n,r}$, which is required to be sufficiently small relative to $\rho$ as $n \rightarrow \infty$. Consequently, the theorem reveals that the rounding parameter $r$ will have to be set smaller when 
\begin{itemize}
\item[(a)] the true function $\eta$ is rougher
\item[(b)] the spline order $m$ is larger
\item[(c)] the predictor distribution $f$ is rougher
\item[(d)] the sample size $n$ is larger.
\end{itemize}
These conclusions derive directly from the requirement that $D_{n,r} $ be sufficiently small relative to $\rho$ as $n \rightarrow \infty$.

\section{Simulation Study}
\label{sim}

\subsection{Design and Analyses}

We conducted a simulation study to demonstrate the benefits of the rounding parameters. As a part of the simulation, we manipulated two conditions: (a)~the function smoothness (8~levels: see Figure~\ref{fig3}), and  (b)~the number of observations (3~levels: $n=1000k$ for $k\in\{100,200,500\}$). Note that the functions are defined such that $J(\eta_{Aj}) < J(\eta_{Ak})$ and $J(\eta_{Bj}) < J(\eta_{Bk})$ for $j<k\in\{1,2,3,4\}$, so the function smoothness is systematically manipulated. We generated $y_{i}$ by (a)~independently sampling the predictor(s) from a uniform distribution, (b)~independently sampling $e_{i}$ from a standard normal distribution, and (c)~defining the observed response as $y_{i}=\eta(\mathbf{x}_{i})+e_{i}$ for $i\in\{1,\ldots,n\}$.

Then, we fit a nonparametric regression model using six different methods: Method~1 is an SSANOVA using unrounded data \citep[see][]{Helwig+Ma:2015}, Method~2 is an SSANOVA with $r=.01$,  Method~3 is an SSANOVA with $r=.02$, Method~4 is an SSANOVA with $r=.05$, Method~5 is standard GAM implemented through \citeapos{mgcv} \texttt{gam.R} function, and Method~6 is batch-processed GAM implemented through \citeapos{mgcv} \texttt{bam.R} function. Methods 1--4 are implemented through \citeapos{bigsplines} \texttt{bigspline.R} function (for $\eta_{Ak}$) and \texttt{bigssa.R} function (for $\eta_{Bk}$). 

For the $\eta_{Ak}$ functions we used $q=21$ knots to fit the model, and for $\eta_{Bk}$ functions we used $q=100$ knots. For Methods~1--4, we used a bin-sampling approach to select knots spread throughout the covariate domain \citep{Helwig+Ma:prep}; for Methods~5 and 6, we used the default \texttt{gam.R} and \texttt{bam.R} knot-selection algorithm \citep[see][]{mgcv}. For each method, we used cubic splines and selected the smoothing parameters that minimized the GCV score. Given the optimal smoothing parameters, we calculated the fitted values, and then defined the true mean-squared-error (MSE) as $(1/n)\sum_{i=1}^{n}(\eta(\mathbf{x}_{i}) - \hat{y}_{i})^{2}$. Finally, we used 100 replications of the above procedure within each cell of the simulation design.

\subsection{Results}

\begin{figure*} 
\centering
 \scalebox{.55}{\includegraphics{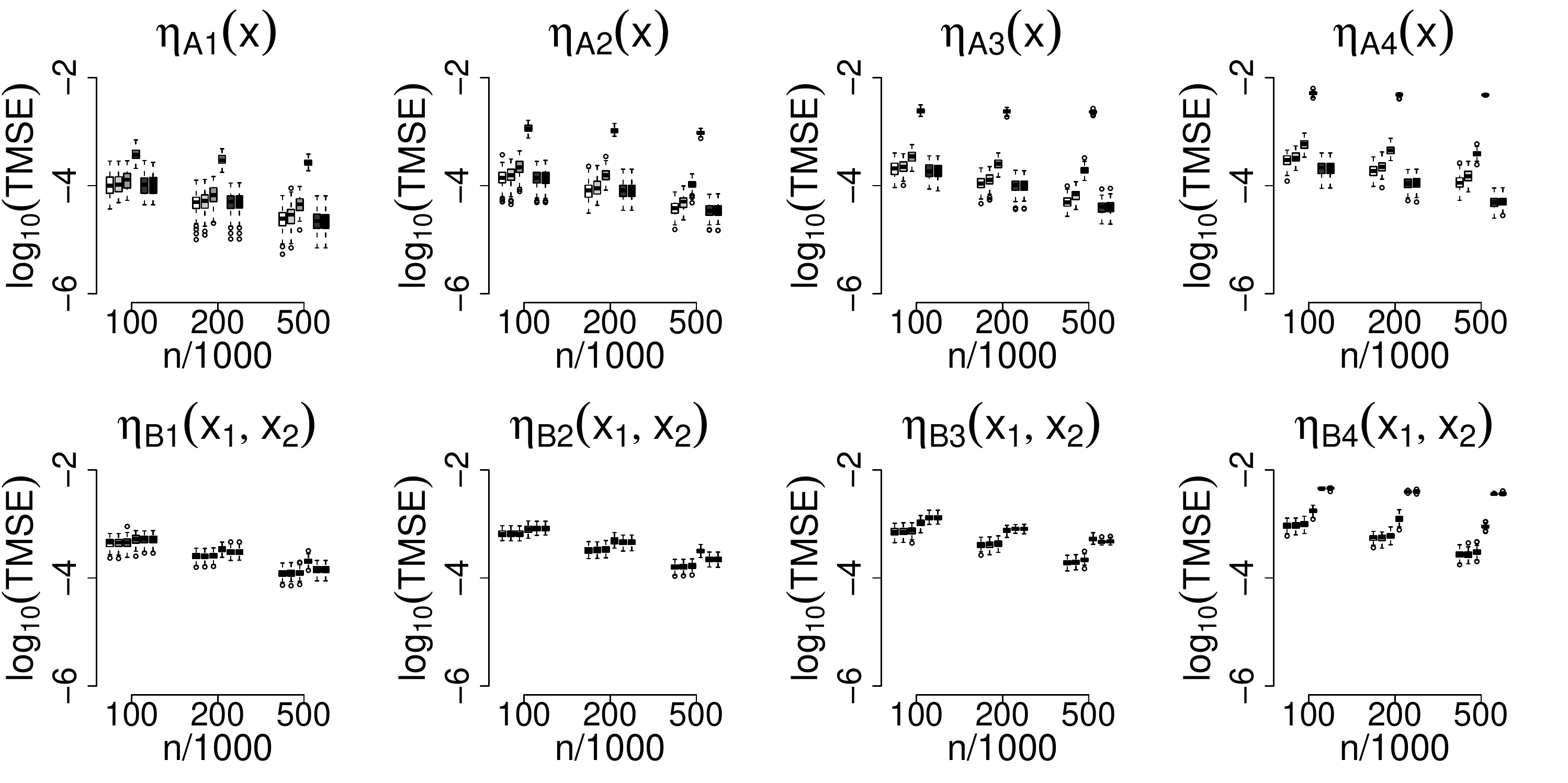}} 
 \caption{\label{fig5} Simulation true MSEs on log-10 scale. Within each sample size, the six boxes correspond to Methods 1--6. Method~1 is SSANOVA with no rounding, Method~2 is SSANOVA with $r=.01$, Method~3 is SSANOVA with $r=.02$, Method~4 is SSANOVA with $r=.05$, Method~5 is \texttt{gam.R}, and Method~6 is \texttt{bam.R}.}
\end{figure*}

The true MSE for each combination of simulation conditions is plotted in Figure~\ref{fig5}. 
First, note that for each method, the true MSE decreased as $n$ increased, which was expected. Next, note that all of the methods recovered $\eta$ quite well (i.e., all MSEs smaller than 0.01). Comparing Methods 1--4, it is evident that setting $r\in\{.01,.02\}$ introduced minimal bias to the resulting solution. In contrast, setting $r=.05$ produced a more noticeable bias, particularly when analyzing the more jagged $\eta_{Ak}$ and $\eta_{Bk}$ functions, i.e., those with larger $k$. However, the bias introduced with $r=.05$ was small relative to the norm of $\eta$, so there is little practical difference between the solutions with $r\in\{.01,.02,.05\}$. Examining the true MSEs of Methods 5 and 6, it is apparent that the standard GAM performed almost identical to the batch-processed GAM throughout the simulation. 

Comparing the true MSEs of Methods 1--4 to those of Methods~5 and 6, it apparent that the SSANOVAs performed similar to the GAMs in every simulation condition. In the one-dimensional case ($\eta_{Ak}$ functions), the GAMs have slightly smaller true MSEs for $k\in\{3,4\}$, but the difference is trivial compared to the norm of the $\eta_{Ak}$ functions. In the two-dimensional case ($\eta_{Bk}$ functions), the SSANOVAs have slightly smaller true MSEs for $k\in\{3,4\}$. Differences between the SSANOVA and GAM solutions are most pronounced when analyzing the $\eta_{B4}$ function; in this case, the median true MSE of the GAM  solutions is over 10 times larger than the corresponding median of the SSANOVA solutions with $r\in\{NA,0.01,0.02\}$. However, the difference is still quite small compared to the norm of the $\eta_{B4}$ function.

The median analysis runtimes (in seconds) for each simulation condition are displayed in Tables~\ref{tab1} and \ref{tab2}. First, note that for each method, the runtime increased as $n$ increased, which was expected. Next, note that the runtimes for Methods~1, 5, and 6 were substantially larger than the corresponding runtimes of Methods~2--4. When analyzing the $\eta_{Ak}$ functions, the median runtimes for Methods 2--4 were less than one-tenth of a second  for all examined $n$, and were anywhere from 40--60 times faster than the median runtimes for Methods 5 and 6. When analyzing the $\eta_{Bk}$ functions, the median runtimes for Methods 3--4 were less than one second for all examined $n$, and were anywhere from 10--20 times faster than the median runtimes for Methods 5 and 6. 

\begin{table*}
\caption{\label{tab1} Median runtimes (seconds) for $\eta_{Ak}$ functions.}
{\footnotesize
\begin{tabular}{|l|rrr|rrr|rrr|rrr|}
\hline
     &  \multicolumn{3}{c}{$\eta_{A1}$} &  \multicolumn{3}{|c}{$\eta_{A2}$} &  \multicolumn{3}{|c}{$\eta_{A3}$} &  \multicolumn{3}{|c|}{$\eta_{A4}$} \\
     &  100  & 200  & 500  & 100   &200  & 500   &100  & 200  & 500   &100   &200  & 500 \\
\hline
Method 1 ($r=$ NA)   &  0.35  &  0.64   &   1.31  &  0.37  &  0.64  &  1.28  &  0.30  &  0.64  &   1.31  &   0.36 &   0.64  &  1.31 \\ 
Method 2 ($r=0.01$)  &  0.02  &  0.03   &   0.07  &  0.02  &  0.03  &  0.07  &  0.02  &  0.03  &   0.07  &  0.02  &  0.03   &  0.07 \\
Method 3 ($r=0.02$)  &  0.02  &  0.03   &   0.06  &  0.02  &  0.03  &  0.06  &  0.01  &  0.03  &   0.06  &  0.01  &  0.03   &  0.06 \\ 
Method 4 ($r=0.05$)  &  0.01  &  0.03   &   0.06  &  0.02  &  0.02  &  0.06  &  0.01  &  0.02  &   0.06  &  0.01  &  0.02   &  0.06 \\
Method 5 (GAM)        &  1.44  &  2.24   &   4.05  &  1.40  &  2.11  &  4.03  &  1.47 &   2.12  &   4.06  &  1.40  &  2.11   &   4.06 \\
Method 6 (BAM)         & 1.35  &  2.02   &   4.26  &  1.37  &  2.05  &  4.30  &  1.32  &  2.05  &   4.28  &  1.38  &  2.05   &  4.29 \\
\hline
\end{tabular}
}
\end{table*}

\begin{table*}
{\footnotesize
\caption{\label{tab2} Median runtimes (seconds) for $\eta_{Bk}$ functions.}
\begin{tabular}{|l|rrr|rrr|rrr|rrr|}
\hline
     &  \multicolumn{3}{c}{$\eta_{B1}$} &  \multicolumn{3}{|c}{$\eta_{B2}$} &  \multicolumn{3}{|c}{$\eta_{B3}$} &  \multicolumn{3}{|c|}{$\eta_{B4}$} \\
     &  100  & 200  & 500  & 100   &200  & 500   &100  & 200  & 500   &100   &200  & 500 \\
\hline
Method 1 ($r=$ NA)    &   3.80   &   6.60   &   14.84  &  3.80   &   6.60   &  14.82   &   3.81 &   6.61  & 14.85   &  3.81  &  6.60  &  14.85 \\
Method 2 ($r=0.01$)   &   0.85   &   0.80   &     1.35  &  0.85   &   0.80   &    1.34   &   0.85  &  0.80  &   1.35   &  0.85  & 0.80   &    1.35 \\
Method 3 ($r=0.02$)   &   0.34   &   0.51   &     0.99  &  0.34   &   0.51   &    0.99   &   0.34  &  0.51  &    0.99  &  0.34  & 0.51   &    0.99 \\ 
Method 4 ($r=0.05$)   &   0.28    &  0.43   &     0.90  &  0.28   &   0.43   &    0.90   &   0.28 &   0.43  &    0.90  &  0.28  & 0.43   &    0.90 \\
Method 5 (GAM)         &    4.48   &   9.16  &    22.31  &  4.45  &   9.12   &   22.29  &   4.45  &  9.16  &  22.38  &  4.50  & 9.20   &  22.43 \\
Method 6 (BAM)         &    4.75   &   7.81  &    18.55  &  4.73  &   7.78   &   18.55  &   4.74  &   7.80  &  18.61  &  4.77  & 7.85   &  18.65 \\
\hline
\end{tabular}
}
\end{table*}

\section{Real Data Example}
\label{ex}

\subsection{Data and Analyses}
\begin{figure} 
\centering
 \scalebox{.75}{\includegraphics{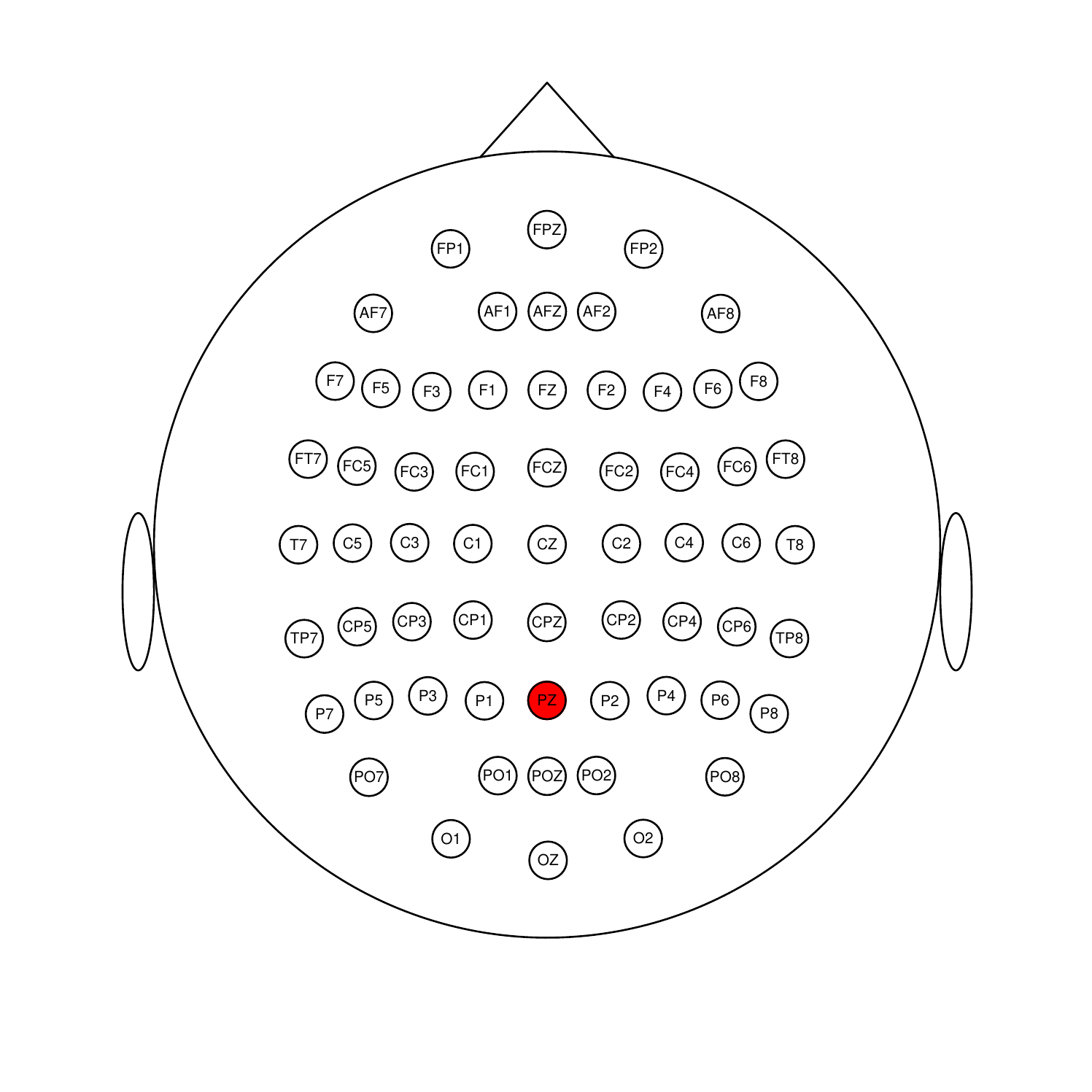}} 
 \caption{\label{fig6} Depiction of the 61-channel EEG cap. The Pz electrode is highlighted in red. Created using the \texttt{eegcap} function in the \texttt{eegkit} R package \citep{eegkit}.}
\end{figure}

To demonstrate the practical benefits of the rounding parameters when working with real data, we use electroencephalography (EEG) data obtained from \citet*{Bache+Lichman:2013}. Note that EEG data consist of electrical activities that are recorded from various electrodes on the scalp, and EEG patterns are used to infer information about mental processing. The EEG data used in this example were recorded from both control and alcoholic subjects participating in an experiment at the Henri Begleiter Neurodynamic Lab at SUNY Brooklyn. The data were recorded during a standard visual stimulus event-related potential (ERP) experiment using a 61-channel EEG cap (see Figure~\ref{fig6}). The data were recorded at a frequency of 256 Hz for one second following the presentation of the visual stimulus.

For the example, we analyzed data from the Pz electrode of 120 subjects (44 controls and 76 alcoholics), and we used 10 replications of the ERP experiment for each subject.\footnote{Note that data from subjects \texttt{co2a0000425} and \texttt{co2c0000391} were excluded from the analysis due to small amounts of data, and we used the first 10 replications for each subject.} This resulted in $n=$ 307,200 data points (120 subjects $\times$ 256 time points $\times$ 10 replications). We analyzed the data using a two-way SSANOVA on the domain $[0,1]\times\{1,2\}$, where the first predictor is the time effect and the second predictor is the group effect (control vs.\ alcoholic); see the Appendix for an explanation of how the rounding parameter can be applied when working with continuous and nominal predictors. We used a cubic spline for the time effect, a nominal spline for the group effect, and $q=50$ bin-sampled knots. Finally, we fit the model both with the unrounded data and with the time covariate rounded to the nearest .01 second (i.e., $r=.01$ on the interval [0,1]); note that setting $r=.01$ for the time covariate results in $u=202$ unique covariate vectors, which is substantially less than the original $n=307200$ data points.

\subsection{Results}

\begin{figure*}
\centering 
 \scalebox{.55}{\includegraphics{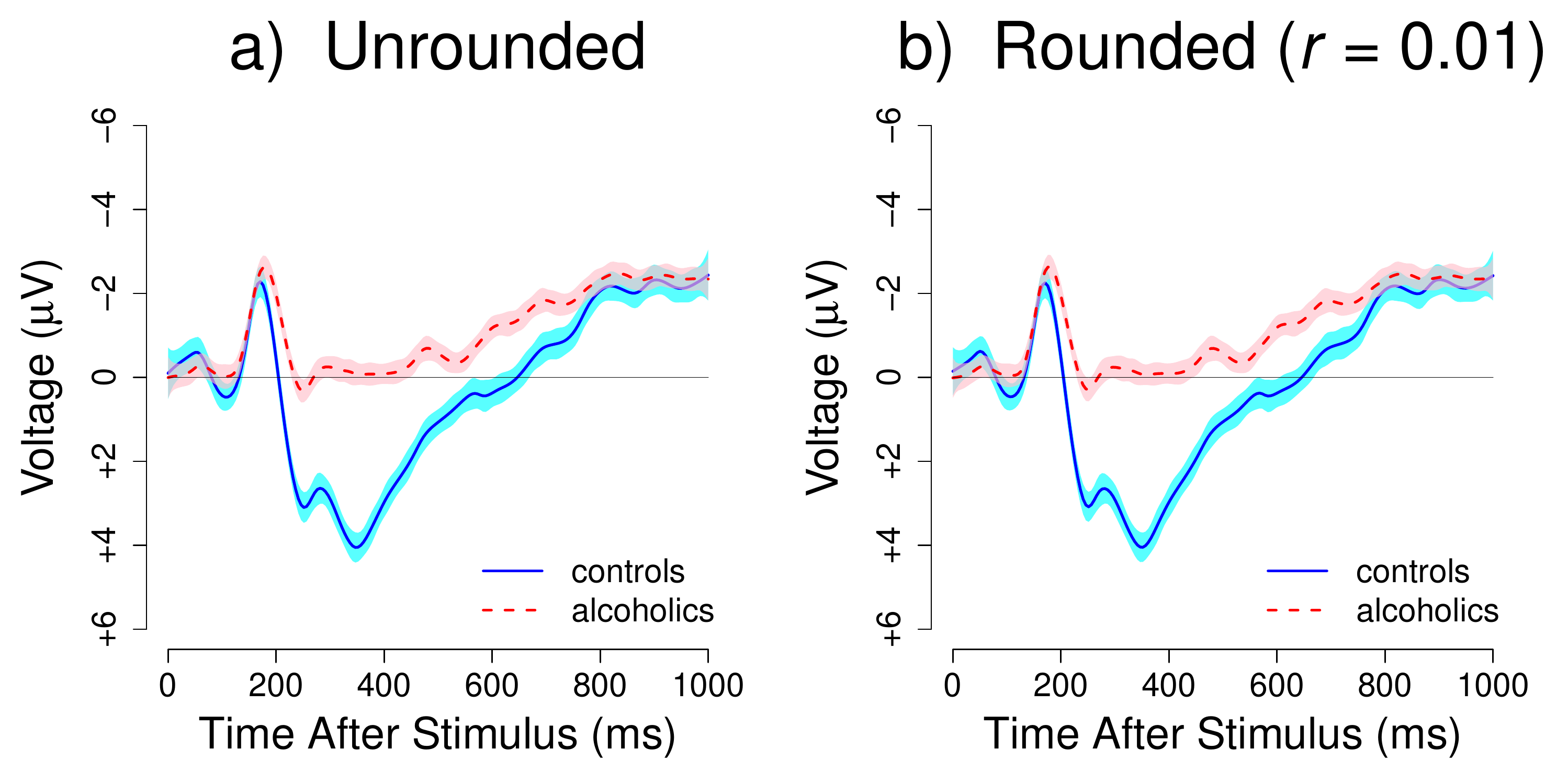}} 
 \caption{\label{fig7} Predicted ERPs using the unrounded data (a) and rounded data (b). Shaded regions give a 99\% Bayesian confidence interval around $\hat{\eta}$. Created using the \texttt{eegtime} function in the \texttt{eegkit} R package \citep{eegkit}.}
\end{figure*}

The predicted ERPs for the unrounded and rounded data are plotted in Figure~\ref{fig7}. 
Note that there are no practical differences between the two solutions (c.f.\ Figure~\ref{fig7}a,b). Furthermore, note that both solutions produced a GCV score of GCV=85.96 and variance-accounted-for value of $R^{2}=0.03$, suggesting that the rounded solution fits the data as well as the unrounded solution. It is also worth noting that the unrounded solution took over five times longer to fit compared to the rounded solution; furthermore, the unrounded solution required a substantial amount of RAM to fit the model, whereas the rounded solution is easily fittable on a standard laptop or tablet.

Comparing the estimated ERPs of the controls and alcoholics, there are obvious differences (see Figure~\ref{fig7}). In particular, the alcoholic subjects are missing the P300 component of the ERP waveform (i.e., large positive peak occurring about 300 ms after the stimulus). Note that the P300 component is thought to relate to a subject's internalization and/or categorization of stimuli, so these results suggest that alcoholic subjects have different information processing patterns for standard visual stimuli. This finding is consistent with previous findings regarding EEG patterns of alcoholic subjects \citep[see][]{Porjesz+EtAl:1980,Porjesz+EtAl:1987}, and some research suggests that this sort of EEG pattern may predispose individuals to alcoholism \citep[see][]{Porjesz+Begleiter:1990a,Porjesz+Begleiter:1990b}.

\section{Discussion}

This paper proposes the use of rounding parameters to overcome the computational burden of fitting nonparametric regression models to super-large samples of data. By rounding each predictor to a given precision (e.g., 0.01), it is possible to estimate $\eta$ using the $u \ll n$ unique rounded predictor variables. We have provided a simple Taylor heuristic that justifies the use of a small rounding parameter (e.g., $r=.01$) when using cubic smoothing splines for $x\in[0,1]$. Furthermore, we have provided methods for assessing the finite sample and asymptotic performance of the rounded SSANOVA estimator in various situations. 

The simulation study and EEG example clearly demonstrate the benefits of the proposed rounding parameters. When fitting nonparametric regression models with large $n$, the simulation results reveal that setting $r_{j}\leq.05$ can result in substantial computational savings without introducing much bias to the solution. Furthermore, the EEG data example reveals that there are no practical differences between the unrounded and rounded solutions (using $r=.01$) when analyzing real data. Thus, the rounding parameters offer a fast and stable method for fitting nonparametric regression models to very large samples.

In addition to providing a fast method for smoothing large datasets, the rounding parameters are also quite memory efficient. Because the rounding approach only uses the unique rounded-covariate values, it is never necessary to construct the full $n\times q$ model design matrix (or the $n\times n$ smoothing matrix). So, using the rounding parameters, it is possible to fit nonparametric regression models to very large samples using a standard laptop or tablet, e.g., all of the rounded SSANOVA models in this paper are easily fittable on a laptop with 4 GB of RAM. As a result, typical researchers now have the ability to discover functional relationships in super-large data sets without needing access to supercomputers or computing clusters.

As a final point, it should be noted that in some cases (e.g., large $p$) the number of unique rounded-covariate values may be very large. In such cases, forming the $u\times q$ model design matrix may require a substantial amount of memory (because $u$ is so large). However, as is noted in \citet*{Helwig:Phd} and \citet*{Helwig+Ma:2015}, fitting an SSANOVA model only depends on various crossproduct vectors and matrices. So, if $u$ is too large to form the full $u\times q$ model design matrix, then the needed crossproduct statistics can be formed in a batch-processing manner similar to the approach used by \citeapos{mgcv} \texttt{bam.R} function. 

\newpage

\section*{Appendix: Rounding Algorithm}

In this section, we provide algorithms for rounding SSANOVA predictors and obtaining the sufficient statistics for the SSANOVA estimation. The first algorithm assumes that all of the covariates are continuous; extensions for nominal covariates will be discussed after the presentation of the initial algorithm.

First, let $r_{j}\in(0,1]$ denote the rounding parameter for the $j$-th predictor, let $\tilde{\mathbf{x}}_{j}$ denote the $n \times 1$ vector containing the $j$-th predictor's scores, and let $x_{(i)j}$ denote the $i$-th order statistic of the $j$-th predictor. Next, initialize $\mathbf{g}\equiv\{1\}_{n\times1}$ and $h\equiv1$, and then calculate
\[
\begin{split}
&\mathrm{for} \ j \in\{1,\ldots,p\}\\
&\qquad 1. \ \ \mathbf{g} \leftarrow \mathbf{g} + h[\mathrm{rd}\{(1/r_{j})(\tilde{\mathbf{x}}_{j}-x_{(1)j})/(x_{(n)j}-x_{(1)j})\}]\\
&\qquad 2. \ \ h \leftarrow \mathrm{rd}(1+1/r_{j})h\\
&\mathrm{end}
\end{split}
\]   
where the rounding function $\mathrm{rd}\{\cdot\}$ rounds the input to the nearest integer. After running the for loop, we have $g_{i}\in\{1,\ldots,u\}$, where $g_{i}$ denotes the $i$-th element of $\mathbf{g}$, and $u$ is the total possible number of unique covariate vectors; thus, the vector $\mathbf{g}$ indexes the multi-dimensional rounded-covariate score for each observation. 

The above result implies that the unique rounded-covariate scores (i.e., $\tilde{\mathbf{z}}_{t}$) can be obtained by sorting the predictors according to the $g_{i}$ values, and then sampling one observation's covariate vector from each unique $g_{i}$ value. Similarly, once the data is sorted according to the $g_{i}$ values, the sum of the response at each unique covariate (i.e., $\tilde{y}_{t}$) and the number of observations at each unique covariate (i.e., $w_{t}$) can be easily calculated. Lastly, after calculating $\|\mathbf{y}\|$, the SSANOVA model can be fit using the sufficient statistics from the rounded solution, i.e.,  $\tilde{\mathbf{z}}_{t}$, $\tilde{y}_{t}$, and $w_{t}$. 

As we previously mentioned, the above algorithm can be modified to include nominal covariates as well. When working with nominal covariates, the algorithm assumes that all nominal covariates are of the form $x_{ij}\in\{1,\ldots,f_{j}\}$ where $f_{j}$ is the number of factor levels of the $j$-th covariate. Assuming that $x_{ij}\in\{1,\ldots,f_{j}\}$, both steps of the rounding algorithm need to be slightly modified: 
\[
\begin{split}
&\mathrm{for} \ j \in\{1,\ldots,p\}\\
&\quad \mbox{If } x_{ij} \mbox{ is continuous} \\
&\qquad 1. \ \ \mathbf{g} \leftarrow \mathbf{g} + h[\mathrm{rd}\{(1/r_{j})(\tilde{\mathbf{x}}_{j}-x_{(1)j})/(x_{(n)j}-x_{(1)j})\}]\\
&\qquad 2. \ \ h \leftarrow \mathrm{rd}(1+1/r_{j})h\\
&\quad \mbox{Else if } x_{ij} \mbox{ is nominal} \\
&\qquad 1. \ \ \mathbf{g} \leftarrow \mathbf{g} + h(\tilde{\mathbf{x}}_{j}-1) \\
&\qquad 2. \ \ h \leftarrow f_{j}h\\
&\mathrm{end}
\end{split}
\]
Using this simple modification, the rounding algorithm can be efficiently applied to any combination of continuous and nominal covariates.

\bibliography{nehref}
\bibliographystyle{mychicago}

\end{document}